\documentclass{appolb}

\usepackage{graphics}
\usepackage{graphicx}
\usepackage{epsfig}
\usepackage{color}
\usepackage{hyperref}
\usepackage{latexsym}
\usepackage{amsmath}
\usepackage{amsthm}
\usepackage{amsfonts}
\usepackage{amssymb}
\usepackage{dsfont}
\usepackage{bm}
\usepackage{graphics,psfrag}
\usepackage{graphicx,psfrag}
\usepackage[tight]{units}

\newcommand{\eq}[1]{(\ref{#1})}
\newcommand{\fig}[1]{Fig.~{\ref{#1}}}

\newcommand{\sect}[1]{Sec.~{\ref{#1}}}

\newcommand{\RDH}[1]{{\mathrm{red}(#1)}}

\newcommand{\LMH}[1]{{\mathrm{lm}(#1)}}

\begin{document}
\title{Effects of the low lying Dirac modes on excited hadrons in lattice QCD%
\thanks{Presented at \emph{Excited QCD 2012}, Peniche, Portugal.}%
}
\author{L.~Ya.~Glozman, C.~B.~Lang, M.~Schr\"ock\thanks{Speaker}
\address{Institut f\"ur Physik, FB Theoretische Physik, Universit\"at
Graz, A--8010 Graz, Austria}
% \\
% {Third Author of different affiliation
% }
% the Name(s) of other Author(s)
% \address{affiliation}
}
\maketitle
\begin{abstract}
Chiral symmetry breaking in Quantum Chromodynamics is associated with the low lying spectral modes of the Dirac operator according to the Banks--Casher relation. Here we study how removal of  a variable number of low lying modes from the valence quark sector affects the masses of the ground states and 
first excited states of baryons and mesons in two flavor lattice QCD.
\end{abstract}
  
\section{Introduction}

In \cite{Lang:2011qy,Lang:2011ai} the effect of removing the lowest Dirac eigenmodes
from valence quark propagators which enter meson correlators has been studied.
It has been found that the exponential decay signal of the pion correlator vanishes whereas
those of the scalar, vector and axial vector currents became essentially
better. Moreover, the removal of the lowest Dirac eigenmodes has been shown to
restore the chiral symmetry in accordance with the Banks--Casher relation: the
masses of the lowest vector and axial vector states became degenerate.
Including more particles and excitations above the ground state is a crucial step
in understanding the destiny of confinement upon artificial chiral
symmetry restoration.
In our recent work \cite{Glozman:2012fj} the $b_1$ meson, the
nucleon and Delta baryons of both parities as well as the first excited states of
most of the hadrons under investigation have been included.
Besides showing the existence of bound states, i.e., confined quarks after having 
artificially restored the chiral symmetry, this work gives insights in the origin
of the $\Delta - N$ mass splitting and its relation to chiral symmetry.

\section{Method}\label{sec:method}
In order to study the effects of Dirac low modes on ground and excited states
of hadrons we modify the valence quark propagators which enter the interpolating
fields of the particles.
To be explicit, we construct \emph{reduced} quark propagators
\begin{equation}\label{eq:red5}
	S_\RDH{k}=S-S_\LMH{k}\equiv S- \sum_{i\le k} \mu_i^{-1} |{v_i}\rangle\langle{v_i}|\gamma_5
\end{equation}
where $S$ is the standard (untruncated) quark propagator
and the low mode part $S_\LMH{k}$ is explicitly constructed from the
eigenvalues $\mu_i$ and eigenvectors $|{v_i}\rangle$ of the Hermitian Dirac operator $\gamma_5 D$. 
The reduction parameter $k$ in \eq{eq:red5} gives the number of
low eigenmodes of the Dirac operator that have been removed from the quark
propagator. We will adopt reduced quark propagators with $k=0, 2, 4, 8, 12,
16, 20, 32, 64, 128$.

These truncated quark propagators can subsequently be used as an ingredient 
for the construction of
hadron interpolators. In this way we can study the
evolution of hadron masses as a function of the number of the subtracted
lowest Dirac eigenmodes. By increasing the number of the subtracted eigenmodes
we gradually remove the chiral condensate of the vacuum and thus
artificially restore the chiral symmetry. 
In \cite{Schrock:2011hq} the latter has been demonstrated on basis of the 
momentum space quark propagator (in Landau gauge) where the dynamically
generated mass of the quark propagator has been shown to disappear as more and more 
Dirac low modes were excluded.

Note that $k$ itself is not
the relevant truncation scale since $k$ has to scale with the lattice volume when keeping
the  physics constant. Instead, we introduce a cutoff parameter $\sigma$
such that all $\mu_k$ with $|\mu_k|<\sigma$ are excluded at reduction level $\sigma$.

\section{The setup}\label{sec:setup}
\subsection{Dirac operator}
For our work we use the so-called chirally improved (CI) Dirac operator
\cite{Gattringer:2000js,Gattringer:2000qu} which
is an approximate solution to the Ginsparg--Wilson equation and
therefore offers better chiral properties compared to the Wilson Dirac operator
while requiring about one order of magnitude less computation time than
the chirally exact overlap operator.

\subsection{Gauge configurations}

Our investigation has been carried out on 161 gauge field configurations 
\cite{Gattringer:2008vj,Engel:2010my} that were generated for two
degenerate dynamical light CI fermions with a corresponding  pion mass
$m_\pi=\unit[322(5)]{MeV}$. The lattice size is $16^3\times 32$ and the
lattice spacing $a=\unit[0.144(1)]{fm}$.

\section{Results}

Here we list the hadrons under study, give details of the calculation
and finally present the evolution of the hadron masses under Dirac low mode truncation. 

\subsection{Mesons}
We restrict ourselves to the study of isovector mesons of spin 1. 
Isoscalars require the inclusion of disconnected diagrams which are
too costly with the CI Dirac operator.
The scalar meson $a_0$ 
and the pseudoscalar  pion under Dirac low mode removal have been investigated in
\cite{Lang:2011qy,Lang:2011ai}.
Therefore, the studied nonexotic channels are the $J^{PC}$ combinations
$1^{--}$ ($\rho$), $1^{++}$ ($a_1$) and $1^{+-}$ ($b_1$). 

The variational method \cite{Luscher:1990ck,Michael:1985ne}
combined with standard interpolating fields and 
three different kinds of quark sources (narrow
and  wide Gaussian shape sources
\cite{Gusken:1989ad,Best:1997qp} and a derivative source)
enable us to extract some information about excited states.

Diagonalisation of the cross-correlation matrix for lattice interpolators with the appropriate
quantum numbers results in eigenvalues with exponential decay $\exp(-E_n t)$.
More details of the calculation can be found in \cite{Glozman:2012fj}
where we also show sample plots for all particles of the eigenvalues and
plots of the eigenvector components and the effective masses including  fit ranges
and values. From the latter the energy values can be determined from exponential fits  to
the eigenvalues.
For the $a_1$ and $b_1$ we could only extract the values of the ground state masses reliably
whereas for the $\rho$ we found a clear signal for the two lowest lying states and consequently
performed fits to these two states.

\begin{figure}
	\center
	\includegraphics[width=1.0\textwidth]{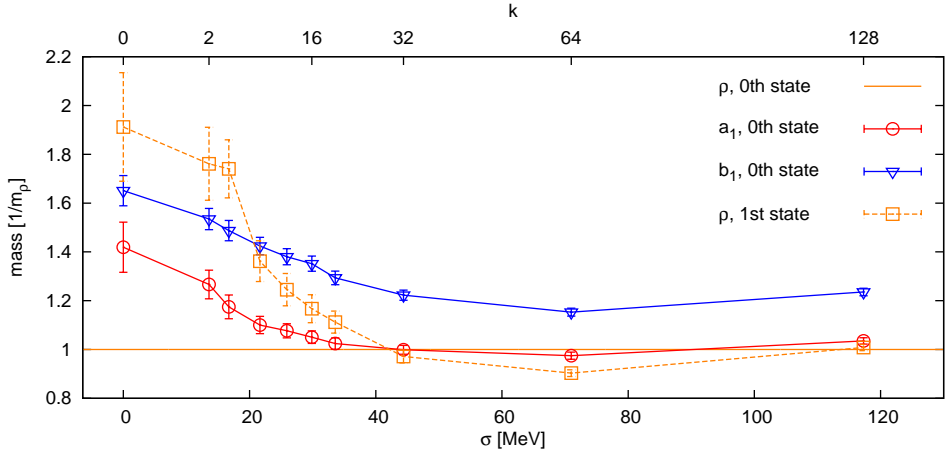}
  \caption{Evolution of the meson masses under Dirac low mode removal.
		Masses in units of the $\rho$-mass at the corresponding truncation level.
}\label{fig:mesons}
\end{figure}

In \fig{fig:mesons} we show the masses of the mesons under study as a function
of the truncation level $\sigma$ (see \sect{sec:method}) in units of the $\rho$-mass
at each truncation level.
We also give the corresponding truncation index $k$ on the
upper abscissa of the plot.

As was already explored in \cite{Lang:2011qy}, the masses of the $\rho$ and $a_1$ approach
each other when subtracting all eigenmodes with a mass smaller or equal to 
approximately two bare quark masses (in our study the bare quark mass is $\approx\unit[15]{MeV}$)
which indicates restoration of the $SU(2)_R \times SU(2)_L$ chiral symmetry.
Furthermore, the additional degeneracy of the $\rho$ and $a_1$ with the $\rho'$ state 
hints to a higher symmetry which has the chiral symmetry as a subgroup.
The fact that the $b_1$ mass remains clearly heavier tells us not only that the $U(1)_A$
symmetry remains broken but also gives strong evidence in the meson sector that we are
still confronted with confined particles instead of two free quarks traveling next
to each other.

\subsection{Baryons}

We analyze the nucleon and $\Delta$ baryons with positive and 
negative parity. For the interpolators we use two different Gaussian smeared quark
sources \cite{Glozman:2012fj}.
For the nucleon we adopt interpolating fields with three different Dirac
structures.
We use parity projection
for all baryons and Rarita--Schwinger projection for the $\Delta$
\cite{Engel:2010my}.
For all details of the calculation we refer again to \cite{Glozman:2012fj}.

\begin{figure}
	\center
	\includegraphics[width=1.0\textwidth]{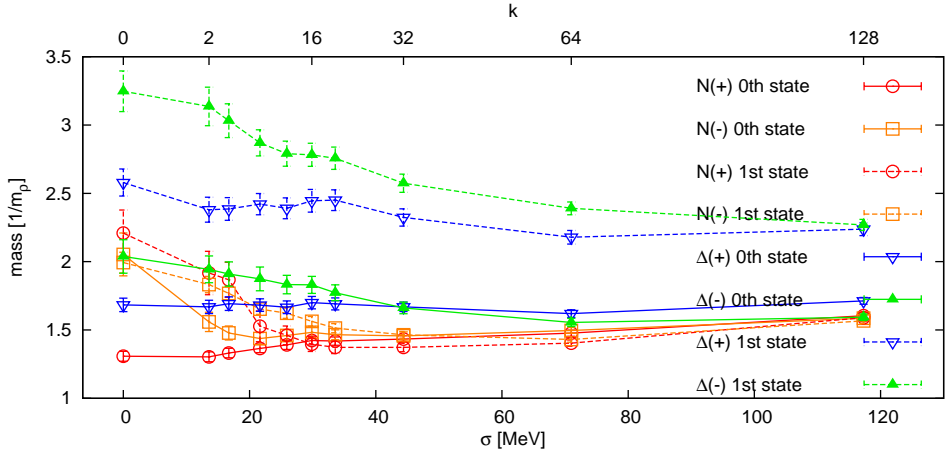}
  \caption{Evolution of the baryon masses under Dirac low mode removal.
		Masses in units of the $\rho$-mass at the corresponding truncation level.
}\label{fig:baryons}
\end{figure}

In \fig{fig:baryons} we show the mass values of the ground and first excited states 
of all baryons as a function of the truncation level.
We find the nucleons of positive and negative parity 
($J^P= \frac{1}{2}^+$ and $J^P= \frac{1}{2}^-$) to become
degenerate upon Dirac low mode removal which reveals them as (would be)
chiral partners.
But not only the ground states of the two nucleons end up having the same
mass, we also see again hints to a higher symmetry which relates the ground
and first excited states of the nucleon of both parities.

On the other hand, the first excitations of the Delta with both parities
($J^P= \frac{3}{2}^+$ and $J^P= \frac{3}{2}^-$) remain clearly higher
than the corresponding ground state masses which again manifests the
existence of confined particle states after having artificially restored
the chiral symmetry.

A last observation is that the $\Delta - N$ splitting gets reduced by 
about a factor 2 once the lowest Dirac modes are excluded.
This allows to draw important conclusions:  
the $\Delta-N$ splitting is usually attributed to the hyperfine
spin-spin interaction between the valence quarks
via either the spin-spin color-magnetic interaction
\cite{De_Rujula:1975ge,Isgur:1978xj} or the flavor-spin interaction
related to the spontaneous chiral symmetry breaking
\cite{Glozman:1995fu}. 
When restoring the chiral symmetry the
effective flavor-spin quark-quark interaction becomes impossible. The
color-magnetic interaction remains. Thus, our result suggests that
both these mechanisms are of equal importance to the
$\Delta-N$ splitting.

\section{Conclusions}

We removed a varying number of the lowest Dirac eigenmodes from the valence
quark sector and constructed hadron correlators out of these modified quark
propagators.

In the meson sector we found clear evidence for the restoration of the chiral
symmetry; the $\rho$ and $a_1$ mesons became degenerate. Moreover, the degeneracy
of the $\rho$ and $\rho'$ states hints to the existence of a higher symmetry.
The mass value of the $b_1$ remains clearly larger which tells us that the
$U(1)_A$ symmetry is still broken. Furthermore, the existence of (distinguished)
particle states after having restored the chiral symmetry shows that confinement
remains intact.

In the baryon sector we studied the nucleon and $\Delta$ with positive and
negative parity. Both nucleon ground states and their first excitations
were found to become degenerate which again hints to a higher symmetry than
simply $SU(2)_R \times SU(2)_L$.
The first excitations of the $\Delta$ remain larger than the corresponding ground
state masses which supports our argument that confinement persists
under the artificial restoration of chiral symmetry.
Lastly, the $\Delta - N$ splitting reduces to roughly half its
value  from which we can conclude that the color-magnetic and the flavor-spin interactions
between valence quarks are of equal importance.

\section*{Acknowledgments}
Support by the Research
Executive  Agency (REA) of the European Union under Grant Agreement 
PITN-GA-2009-238353 (ITN STRONGnet) 
and by the Austrian Science Fund (FWF) through grant DK W1203-N16
is gratefully acknowledged.

%\clearpage

%--------------
% Bibliography:
%--------------
\providecommand{\href}[2]{#2}\begingroup\raggedright\endgroup

\end{document}